\documentclass[11pt]{article}

\usepackage[]{acl}
\usepackage{multirow}
\usepackage{times}
\usepackage{latexsym}
\usepackage{amsmath}
\usepackage{rotating}
\usepackage{array}
\usepackage{longtable}
\usepackage[T1]{fontenc}

\usepackage[utf8]{inputenc}

\usepackage{microtype}

\usepackage{inconsolata}

\usepackage{graphicx}

\title{Tracking Articulatory Dynamics in Speech with a Fixed-Weight BiLSTM-CNN Architecture}

\author{
 \textbf{Leena G Pillai\textsuperscript{1,2}}\\
 \textbf{D. Muhammad Noorul Mubarak\textsuperscript{1}}
 \textbf{Elizabeth Sherly\textsuperscript{2}}\\
 \textsuperscript{1}University of Kerala\\
 \textsuperscript{2}Digital University Kerala
\\
 \small{
   \textbf{Correspondence:} \href{leenabelieve@gmail.com}{leenabelieve@gmail.com}
 }
}
\begin{document}
\maketitle
\begin{abstract}
Speech production is a complex sequential process which involve the coordination of various articulatory features. Among them tongue being a highly versatile active articulator responsible for shaping airflow to produce targeted speech sounds that are intellectual, clear, and distinct. This paper presents a novel approach for predicting tongue and lip articulatory features involved in a given speech acoustics using a stacked Bidirectional Long Short-Term Memory (BiLSTM) architecture, combined with a one-dimensional Convolutional Neural Network (CNN) for post-processing with fixed weights initialization. The proposed network is trained with two datasets consisting of simultaneously recorded speech and Electromagnetic Articulography (EMA) datasets, each introducing variations in terms of geographical origin, linguistic characteristics, phonetic diversity, and recording equipment. The performance of the model is assessed in Speaker Dependent (SD), Speaker Independent (SI), corpus dependent (CD) and cross corpus (CC) modes. Experimental results indicate that the proposed model with fixed weights approach outperformed the adaptive weights initialization with in relatively minimal number of training epochs.  These findings contribute to the development of robust and efficient models for articulatory feature prediction, paving the way for advancements in speech production research and applications.  

\end{abstract}
\small{\textbf{keyword}
Acoustic-to-Articulatory Inversion, Smoothing techniques, Articulatory features, Weight initialization, Bidirectional Long Short-Term Memory (BiLSTM)}

\section{Introduction}
Acoustic features play a crucial role as the primary extracted information from speech signals. However, they are the outcomes of the underlying articulatory process. Acoustic-to-Articulatory Inversion (AAI) is a specific research area that focuses on the inverse mapping from acoustic features to their corresponding source articulatory features \cite{1}. In simpler terms, AAI aims to identify or estimate ''how a speaker produces specific speech sounds?''. The relationship between acoustic and articulatory features is inherently one-to-many, and the mapping between them is nonlinear in nature \cite{2}. The advancement of AAI has the potential to enhance various speech technologies, including speech analysis and synthesis \cite{3, 4}, speech and speaker recognition \cite{5, 6, 7, 8}, evaluation of pathological subjects \cite{9}, and the diagnosis of depression from speech \cite{10}.

Articulatory movement tracking is essential for understanding the intricacies of speech production. The most common technologies utilized for tracking articulatory movements, including Magnetic Articulography Methods (e.g., EMA, PMA), Palatograph Methods (e.g., EPG, OPG), and Imaging methods (e.g., MRI, UTI, X-ray) \cite{11, 12, 13, 14, 15, 16}. All these technologies typically require physical contact with the subject. EMA has gained significant prominence in the fields of linguistics, pathological subject analysis and the researches in AAI, due to its advanced sensor-tracking technology. Its effectiveness in capturing and analyzing articulatory movements makes it particularly suitable for research focused on studying specific articulatory aspects.  AAI mapping using EMA generated articulatory features serves as a bridge between the acoustic properties of speech and the underlying articulatory mechanisms, contributing to various fields such as linguistics, speech technology, clinical research, and language learning. Therefore, the EMA sensor dataset was chosen as the preferred technology for tracking and analyzing tongue and lip movements.

Speech is a result of intricate coordination of active and passive articulators within the oral cavity. Estimating the articulatory involvement in speech production remains a crucial area of research. The tongue is the most versatile articulator in speech production but it is not directly visible during speech production as well. This work focuses on estimating the place and manner of the tongue and lip during speech production. A stacked Bidirectional Long Short-Term Memory (BiLSTM) model is employed to map articulatory features from acoustic features. Since articulatory features exhibit smooth sequential patterns, a Convolutional Neural Network (CNN) is employed to enhance the predicted trajectories with fixed weight initialization. By combining the strengths of BiLSTM and CNN, this work aims to improve the understanding of tongue and lip movement dynamics during speech production.

The primary objectives of this work are:
\begin{itemize}
    \item To develop a robust and reliable methodology to estimate the articulatory movements of the tongue and lip from the acoustic speech signal by applying advanced sequential architecture, such as the stacked BiLSTM.
    \item To enhance and smooth the estimated articulatory trajectories using a Convolutional Neural Network (CNN) with fixed weight. 
\end{itemize}
	 
By attaining these goals, the effort intends to extend knowledge of tongue and lip articulatory movements, improve the precision of articulatory tracking, and contribute to the creation of novel tools for the detection and treatment of speech disorders.

\section{Literature Review}
Acoustic to Articulatory Inversion (AAI) aims to understand the connection between the acoustic characteristics of speech and the associated articulatory movements involved in it. Due to the scarcity of parallel corpora, researchers explored statistical techniques to establish relationships between articulatory and acoustic space, including the codebook - based approach \cite{17, 18}. These early techniques, nevertheless, had trouble capturing the intricate and non-linear mapping between the two domains.

Dynamic constraints were introduced to get around these limitations, which significantly improved AAI performance \cite{19}. Researchers subsequently suggested various modelling techniques, such as feedforward neural networks \cite{20}, Gaussian Mixture Models (GMM) \cite{21}, Hidden Markov Models (HMM) \cite{22}, and Mixture Density Networks (MDN) \cite{23, 24}. Even though these methods shown better performance, their architecture exclusively depended on parallel recordings of acoustic-articulatory data, which limits their ability to generalize.

With the increased availability of acoustic-articulatory corpora, Deep Learning (DL) methods have gained popularity in addressing the AAI problem \cite{25, 26, 27}. Despite these advancements, AAI still faces challenges, particularly in subject dependent scenarios with substantial parallel data requirements. To address this, \cite{28} proposes a subject-adaptive AAI method (SA-AAI) using BiLSTM, achieving comparable performance to subject-dependent AAI (SD-AAI) with significantly less training data (\(\sim \)62.5\%), offering a more cost-effective and practical solution. Additionally, \cite{29} addresses the challenges of speech inversion with speaker variability by introducing a vocal tract length normalization (VTLN) technique to minimize speaker-specific details. The proposed VTLN approach transforms acoustic features of different speakers to a target speaker's space, resulting in an 8.15\% relative improvement in the correlation between actual and estimated tract-variable (TV) trajectories, indicating the effectiveness of VTLN in achieving a more robust and accurate speaker-independent speech inversion system.

A novel AAI neural architecture with 1D convolutional filters for spectral sub-band processing, outperformed traditional models in both speaker-dependent and speaker-independent tasks. Further, a lightweight TCN-based system also outperforms BiLSTM for articulatory feature extraction \cite{31}. In \cite{32}, the researchers investigate the application of FastSpeech, a transformer framework that utilizes explicit duration modeling, to achieve consistent alignment between the articulatory movements and phonemes in Phoneme-to-Articulatory (PTA) motion estimation. By leveraging FastSpeech, the researchers aim to improve the performance of phoneme-to-articulatory motion estimation and enhance the understanding of the relationship between speech sounds and articulatory movements. Additionally, \cite{33} explores the use of pretrained Self-Supervised Learning (SSL) features with transformer neural networks for AAI. Instead of traditional MFCCs, SSL features like TERA and DeCoAR, known for their effectiveness in other speech tasks, are evaluated for AAI. The results demonstrate that these SSL features perform well in AAI, achieving subject-specific correlation coefficient (CC) scores similar to the best fine-tuned MFCCs, indicating the promising potential of utilizing SSL features and transformer models in the AAI task. AAI researches are extended to disordered speech as well. In \cite{34}, a set of pretrained SSL features find suitable for articulatory inversion for dysarthric patients. 

The literature highlights that ElectroMagnetic Articulography (EMA) is an effective method for closely examining specific articulatory movements, but its intrusive nature can disrupt natural speech production due to sensor placement on speech organs. In contrast, AAI offers a non-invasive and natural approach to estimate articulatory movements solely from speech recordings. By combining EMA data with acoustic features, a more comprehensive understanding of speech production can be achieved, with EMA capturing precise articulatory movements and acoustic features conveying speech sounds and prosody information. Models trained on this combined feature set from diverse data can generalize to new speakers, enhancing their versatility in real-world applications. The proposed automatic tongue and lip tracking system in this work demonstrates the estimation of articulatory features solely from speech acoustics, using a BiLSTM-CNN architecture, eliminating the need for physical devices.


\section{Proposed Architecture}
 
\begin{figure*}[!t]
    \centering
    \includegraphics[width=1\textwidth]{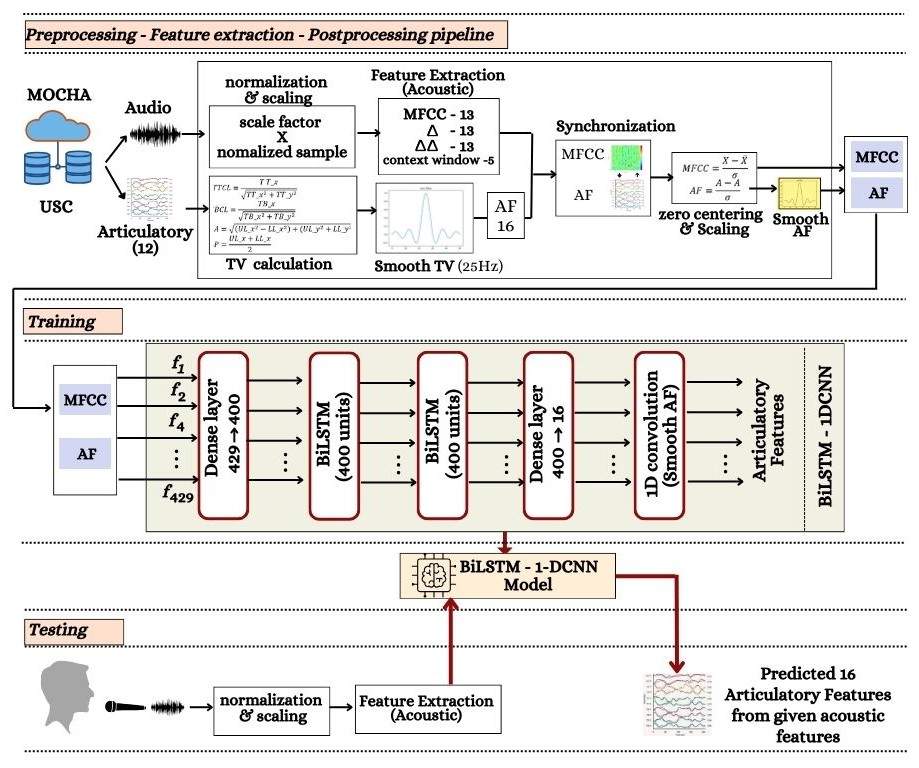}
    \caption{Proposed BiLSTM-CNN Architecture for AAI}
    \label{fig:archi}
\end{figure*}

The proposed architecture seeks to unravel the intricate relationship between the acoustic features of speech and the corresponding movements of speech articulators. The proposed multi-layer approach includes dense transformations, BiLSTM layers for temporal modeling, and a convolutional smoothing network. Together, these components form a stacked ensemble capable of not only modeling the complex temporal dynamics of speech but also refining the predicted articulatory trajectories to be both coherent and continuous. The proposed architecture is illustrated in Fig. \ref{fig:archi}.  

\section{Dataset}
The designed network architecture trained and experimented with publicly available datasets, namely MOCHA \cite{35} and USC-TIMIT \cite{36}, to design their network architecture. The MOCHA corpus, created by Carstens Articulography, included four UK English speakers (two male and two female), while the USC-TIMIT corpus, provided by NDI (Northern Digital Instruments) Wave Speech Research System, comprised three native American speakers (one male and two female). 

Each speaker contributed 460 English sentences recorded at a sampling rate of 16 kHz, resulting in a total of 3220 (460 $*$ 7) English sentences used for experimentation. The EMA system recorded precise x, y coordinates of essential articulatory organs located in the midsagittal plane, encompassing six key areas: Lower Lip (LL\textsubscript x, LL\textsubscript y), Upper Lip (UL\textsubscript x, UL\textsubscript y), Lower Incisor (LI\textsubscript x, LI\textsubscript y), Tongue Tip (TT\textsubscript x, TT\textsubscript y), Tongue Blade (TB\textsubscript x, TB\textsubscript y), and Tongue Dorsum (TD\textsubscript x, TD\textsubscript y). These six articulatory features provided valuable information about the spatial dynamics of speech articulation. 

The study made use of two distinct corpora which were created using different technologies and included speakers from diverse backgrounds. The inclusion of different native speakers in the datasets provided a wide range of linguistic and articulatory characteristics. Moreover, the two corpora utilized different EMA configured devices, which contributed to the dataset's diversity and comprehensiveness, enabling a thorough evaluation of the network's performance. 

\begin{figure}[!h]
    \centering
    \includegraphics[width=1\linewidth]{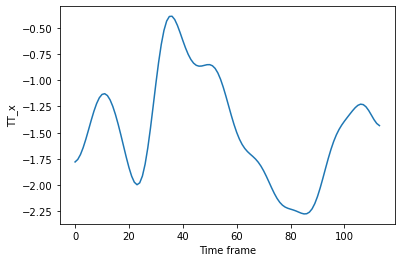}
    \caption{$TT_x$ EMA sensor value of a US speaker for uttering "This was easy for us" }
    \label{fig:USema}
\end{figure}

\begin{figure}[!h]
    \centering
    \includegraphics[width=1\linewidth]{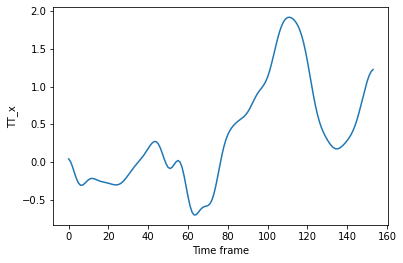}
    \caption{$TT_x$ EMA sensor value of a UK speaker for uttering "This was easy for us"}
    \label{fig:UKema}
\end{figure}

In figure \ref{fig:USema} and \ref{fig:UKema}, illustrated the horizontal movement of tongue tip  ($TT_x$) for uttering the sentence "This was easy for us" by US speaker and UK speaker, respectively. 

The duration of utterance can vary depending of many factors including emphasis, rate of speech, intonation and individual accent. From the corpus it has been noticed that the US speaker might take around 1 to 1.4 seconds to articulate the sentence "This was easy for us" whereas the UK speaker might take relatively long duration, around 1.5 to 2 seconds for the same articulation. Generally, vowel sound in "was" might be pronounced with a shorter vowel sound, closer to "wuhz" by the UK speakers. The US speakers often have the tendency to stretch the vowel sound in "was," making it closer to "waaas". The UK speakers usually drop or less pronouns the 'r' sound at end of a word, therefore "for us" might more sound like "foh us". Whereas US speakers tend to emphasis the 'r' sound, "forr us". The US speaking style is more expressive than UK speakers, they might give stress to different words to convey the important of the context. These highlighted variations are visible throughout the USC and MOCHA corpus.   

This deliberate corpus selection allows to conduct experiments on speaker-dependent, speaker-independent, corpus dependent and cross corpus  scenarios, effectively assessing the performance of the network. 

\section{Preprocessing, Feature Extraction \& Postprocessing}
Each recording system is characterized by its unique recording parameters \cite{37}. This inherent variability in recording parameters presents a noteworthy challenge when amalgamating multiple datasets. Due to the two different geographical origin, the speakers of one dataset carries diverse acoustic and articulatory features. An additional challenge lies in the necessity to synchronize the acoustic features with their corresponding motion trajectory data. To effectively address these challenges, it becomes imperative to conduct preprocessing both prior to and subsequent to feature extraction. This strategic preprocessing ensures data harmonization, extraction precision, and the seamless integration of diverse datasets.

The audio data is initially normalized to ensure that the signal falls within a standardized amplitude range of -1 to 1. Additionally, the audio is rescaled by a factor of 0.5 to facilitate further processing and prevent distortion. The audio data, sampled at a rate of 16 kHz and is divided into frames with a duration of 25 milliseconds each. These frames overlap with a hop time of 10 milliseconds, ensuring comprehensive coverage of the audio signal. From each frame, 13 Mel-Frequency Cepstral Coefficients (MFCCs), 13 first-order delta features (delta coefficients), and 13 second-order delta features (delta-delta coefficients) are calculated. These features are extracted within a context window of 5 frames, which allows for the capture of temporal dependencies and variations in the audio signal effectively. In essence, each frame encompasses information from 5 preceding frames (the past), the current frame itself, and 5 subsequent frames (the future), resulting in a total of 11 frames effectively considered. Consequently, each acoustic feature vector for a frame consists of 429 values (11 frames $*$ 39 features), providing a comprehensive representation of the audio context.

Twelve-dimensional articulatory features were extracted directly from the EMA sensor datasets. Each dimension corresponds to the x and y coordinates of individual sensors, capturing horizontal (x) and vertical (y) trajectory movements during the speech. This detailed articulatory motion representation encompasses six sensor positions, offering comprehensive insights into speech articulation. Within this context, two Tract Variables (TV) gain prominence for analyzing tongue movement \cite{29, 39}. The Tongue Tip Construction Location (TTCL, Eq. (\ref{1})) characterizes tongue tip spatial coordinates, elucidating forward movement, while the Tongue Body Constriction Location (TBCL, Eq. (\ref{2})) quantifies angular constriction formed by Tongue Tip (TT) and Tongue Body (TB), providing insights into tongue coordination \cite{31}.

\begin{equation}\label{1}
   TTCL = \frac{TT_x}{\sqrt{{TT_x}^2 + {TT_y}^2}}
\end{equation}
\begin{equation}\label{2}
   TBCL = \frac{TB_x}{\sqrt{{TB_x}^2 + {TB_y}^2}}
\end{equation}

Additionally, Lip Aperture (LA, Eq. (\ref{3})) and Lip Protrusion (LP, Eq. (\ref{4})) features highlight lip movement dynamics. These extracted features facilitate a comprehensive understanding of articulatory behavior, thereby serving as a foundation for deeper speech analysis and interpretation.

\begin{equation}\label{3}
   LA = \sqrt{\left(UL_x^2 - LL_x^2\right) + (UL_y^2 + LL_y^2)}
\end{equation}
\begin{equation}\label{4}
   LP = \frac{UL_x + LL_x}{2}
\end{equation}

Smoothing applied to the estimated tract variables that can ensure to maintain a smooth and continuous trajectory over time. In this work, the EMA preprocessing pipeline incorporates a windowed Sinc Filter, a key element in optimizing the accuracy of trajectory data. It’s gradual roll-off helps preserve the fundamental shape and features of the trajectory while filtering out high-frequency noise. This filter is entitled to selectively allow frequencies below a designated threshold frequency or cutoff frequency to pass through, effectively eliminating high-frequency noise and distortion that might obscure the desired signal components. In this work, the cutoff frequency (fc) is set at 25 Hz \cite{38}. The application of sinc filter is expressed in Eq. (\ref{5})
\begin{equation}\label{5}
  h = \text{sinc}\left(2 \cdot fc \left(n[i] - \frac{N - 1}{2}\right)\right)
\end{equation}

Here, n[i] represents the index of the current sample, and N is 50 that signifies the length of the filter kernel. However, the practical application of the Sinc filter introduces the challenge of spectral leakage in speech signals. Hanning window is used to address this, Eq. (\ref{6}):
\begin{equation}\label{6}
  \omega = 0.5\left(1 - \cos\left(2\pi \frac{n[i]}{N - 1}\right)\right)
\end{equation}

The ultimate level of finesse is attained through the fusion of the Sinc filter and the Hanning window. This fusion results in a Windowed Sinc filter, represented by Eq.(\ref{7})):
\begin{equation}\label{7}
  W_{sf} = h[i] \cdot \omega[i]
\end{equation}

Following the feature extraction, a synchronization process is applied in which the EMA trajectories, which represent the positions of articulatory organs, are resampled to match the temporal resolution of the MFCC, facilitating a one-to-one correspondence between the two sets of features. Subsequently, the extracted features undergo an Z-score normalization process to standardize their scales and distributions. Normalization may not address irregularities, therefore smoothing applied in post-normalization articulatory data for maintain temporal consistency, and facilitate alignment with acoustic data. After postprocessing, the 39 acoustic features and its corresponding 16 articulatory features are ready to use input for modeling.

\section{Experimental Setup}
\subsection{Network Architecture}
Dense Layer: A fully connected linear layer with 400 units used as an initial layer to extract relevant features from 429 input dimensions (acoustic features), and the Rectified linear activation function (ReLU) is used as an activation function. After applying context window, the current frame x\textsubscript t is a series of $x_t - 5$, $x_t - 4$, ..., $x_t - 1$, $x_t$, $x_t + 1$, $x_t + 2$, ..., $x_t + 5$. The acoustic feature of a single frame x\textsubscript t is 39 and by including the context window the input dimension is 429 (39*11 frames).

BiLSTM: The Long Short-Term Memory is a proven state-of-art architecture for the modeling of time series data \cite{40,41}. In a sequential speech, one input sequence $x_t$ at $t^{th}$ time frame is depend on the previous input sequence $x_{t-1}$ at $t-1^{th}$, and the next input sequence $x_{t+1}$ at $t+1^{th}$ time frame. Therefore, this work implements BiLSTM to learn the dynamic property of the training data through the forward and backward direction. Three gates are encapsulated in LSTM that enables to store the previous relevant information in long term memory, and overcome the vanishing gradient problem. The hidden state (h) holds the short-term memory and the cell state (c) holds the long-term memory (Table \ref{tab:LSTMforbac}).

\begin{table*}[h!]
    \caption{LSTM gate activities in forward and backward direction. Where, w and b are the weight and bias of each corresponding state.}
    \centering
    \begin{tabular}{c|c|c}
        \hline
       \textbf{Gates} & \textbf{Forward} & \textbf{Backward} \\ \hline
       Forget Gate & $f_t={\sigma(w}_f\left(h_{t-1},\ x_t\right)+b_f $ & $f_t={(w}_f\left(h_{t+1},\ x_t\right)+b_f$ \\ &
       $c_t^f=c_{t-1}\ast f_t$ & $c_t^f=c_{t+1}\ast f_t$ \\ \hline
       Input gate & $i_t={(w}_i\left(h_{t-1},\ x_t\right)+b_i$ & $i_t={(w}_i\left(h_{t+1},\ x_t\right)+b_i$\\ & $c_t\acute=tanh(w_c\left(h_{t-1},\ x_t\right)+b_c)$ & $c_t\acute=tanh(w_c\left(h_{t+1},\ x_t\right)+b_c)$ \\ &
       $c_t^i=c_t\acute\ast i_t$ & $c_t^i=c_t\acute\ast i_t$ \\ & $c_t=c_t^f+c_t^i$ & $c_t=c_t^f+c_t^i$\\ 
       \hline 
       Output gate & $O_t={\sigma(w}_o\left(h_{t-1},\ x_t\right)+b_o$ & $O_t={\sigma(w}_o\left(h_{t+1},\ x_t\right)+b_o$\\ & $h_t=o_t\ast tanh(c_t)$ & $h_t=o_t\ast tanh(c_t)$\\
       \hline
    \end{tabular}
    \label{tab:LSTMforbac}
\end{table*}

A BiLSTM is the combination of forward LSTM and the backward LSTM $(BiLSTM = \overrightarrow{LSTM} + \overleftarrow{LSTM})$. To harness the power of LSTMs, two Bi-directional LSTM (BiLSTM) layers are stacked. Each BiLSTM layer comprises 400 units, providing ample capacity to capture intricate temporal dependencies in the acoustic data. Additionally, these BiLSTMs operate in both forward and backward directions, enhancing their ability to learn from both past and future context. This dual-directional learning is especially beneficial for tasks involving sequential data, like AAI, as it enables the network to consider information from all temporal perspectives.

Dense Layer: Following the LSTM layers, the network employs a dense linear layer with the primary purpose of dimensionality reduction. The output from the LSTM layers is 400-dimensional, yet the ultimate objective is to predict 16 articulatory features. Hence, this dense layer acts as a dimensionality transformer, effectively reducing the dimensionality from 400 to the desired 16.

1Dconvolution: During speech production, articulatory movements exhibit a smooth and sequential nature. However, the predicted articulatory features may contain jagged values that do not accurately represent real articulatory movement. In this context, it is more suitable to employ an adaptive smoothing mechanism rather than relying on a predefined cutoff frequency for the smoothing process. To achieve this, a one-dimensional Convolutional Neural Network (CNN) is utilized to dynamically identify the optimal smoothing parameter through learning. The smoothing operation is carried out using a Windowed-Sinc Low-Pass Filter, which is implemented within the Convolutional and pooling layer with a stride value of one. This filter effectively reduces noise and ensures that the predicted articulatory features maintain a smoother and more realistic trajectory. The convolution layer provides smoothed representations of the 16 predicted articulatory features. 

\subsection{Hyperparameter tunning}
Early stopping: In this study, a total of 50 training epochs were initially set as the maximum number of epochs for model training. However, an early stopping mechanism with a patience of 7 was also implemented. This early stopping criterion monitored the model's performance on a validation dataset and halted training if there was no improvement in validation loss for seven consecutive epochs. Interestingly, during experimentation, it was observed that the early stopping criterion often came into effect around the 30th epoch. This suggests that the model achieved a satisfactory level of performance relatively early in the training process, and further training epochs did not significantly contribute to improved results. This efficient use of training epochs helped prevent overfitting and resulted in a more optimized and effective model.

Batch size: The experimentation involved the systematic exploration of different batch sizes, specifically 8, 16, and 32, during the training of the neural network model. Batch size is a critical hyperparameter in BiLSTM that determines the number of data samples used in each iteration of training. Varying the batch size allows to assess its impact on the model's performance and training dynamics. It was observed that a batch size of 8 yielded the most promising results. This choice of batch size was made after rigorous experimentation and evaluation, and it emerged as the configuration that contributed to the best articulatory feature predictions.

Weight Initialization: In this study, two distinct weight initialization strategies were investigated for the convolutional layer responsible for smoothing articulatory features. The first approach, referred to as Adaptive Weight Initialization, entailed initializing the convolutional layer's weights randomly, allowing the model to learn the optimal smoothing characteristics from scratch during training. In contrast, the Fixed Weight Initialization strategy involved keeping the best-performing weight in the low-pass filter frozen and using it as the initial weight for the 1D convolutional layer, thus maintaining constant filter weights throughout training. These experiments aimed to assess whether initializing the convolutional layer with fixed, well-performing weights improved the model's performance compared to allowing the model to learn the filter from random initialization, providing valuable insights into weight initialization's impact on the accuracy and training convergence of the model in the context of articulatory feature smoothing.

\subsection{Dataset preperation}
The total dataset was partitioned into three subsets: 70\% for training, 10\% for validation, and 20\% for testing, to facilitate the evaluation of the proposed model's performance. In the experimental setup, four distinct modes were explored for training and testing the proposed model.

\begin{table}[ht]
  \caption{Dataset preperared for experiment}
  \centering
  \begin{tabular}{|c|c|c|}
    \hline
    \small{\textbf{Mode}} & \small{\textbf{Training (70\%)}} & \small{\textbf{Test (20\%)}} \\
    \hline
    \small{Speaker} & \small{Speaker 1} & \small{Speaker 1} \\
    \small{dependent (SD)} &  & \\
    \hline
    \small{Speaker } & \small {Speaker 1}, ..., n-1 & \small{Speaker n} \\
     \small{independent (SI)} & & \\
    \hline
    \small{Corpus} & \small{MOCHA} & \small{MOCHA} \\
    \small{dependent (CD)} & \small{USC} & \small{USC} \\
    \hline
    \small{Cross Corpus (CC)} & \small{MOCHA} & \small{USC} \\
    & \small{USC} & \small{MOCHA} \\
    \hline
  \end{tabular}
   \label{tab:data}
\end{table}

The experimental setup encompassed a diverse range of testing scenarios to thoroughly assess the performance and adaptability of the proposed neural network model for AAI. The results obtained from these experiments demonstrated that the proposed model exhibited promising performance in predicting articulatory features from acoustic input. Moreover, that it showcased its ability to generalize well across different speakers. The training and test configuration listed in Table \ref{tab:data}. 

\section{Evaluation Parameters}
Two metrics are used to evaluate the performance of the proposed architecture - Pearson Correlation Coefficient (PCC) and Root Mean Square Error (RMSE). 
The Pearson correlation coefficient approach evaluates the strength and direction of the relationship between the estimated and the target articulatory features. Suppose $A = \left\{a_1, a_2, a_3, \ldots, a_n\right\}$ are the target articulatory feature set and $Y=\left\{y_1,\ y_2,\ y_3,\ldots,\ y_n\right\}$ are the predicted articulatory features then the PCC can be defined as Eq (\ref{8}):
\begin{equation}\label{8}
   PCC=\frac{\sum_{i=1}^{n}{(y_i-\bar{y})(a_i-\bar{a})}}{\sum_{i=1}^{n}{\left(y_i-\bar{y}\right)^2{(a_i-\bar{a})}^2}}
\end{equation}

Where $\bar{y}$ and $\bar{a}$ are the mean value of the predicted and actual articulatory values, respectively. The PCC value ranges from 1 to -1. The PCC value of 1 describes an exactly positive correlation, and the value of -1 describes an extreme negative correlation. 
The Root Mean Square Error is used to evaluate the performance of the designed regression model. The proposed architecture is evaluated by measuring the distance between the predicted and the target articulatory features the system is expected to achieve.  
Let’s consider the A and Y feature set described above, the RMSE can be calculated as in Eq. (\ref{9}):
\begin{equation}\label{9}
   RMSE=\sqrt{\frac{\sum_{i=1}^{n}\left(a_i-y_i\right)^2}{N}}
\end{equation}

Where, N is the number of observations. The value of RMSE can be range from 0 to $ \infty $. The value 0 indicate that the error rate is zero.

\section{Result and Discussion}
This section presents a comprehensive analysis of the outcomes obtained from the proposed BiLSTM-CNN model for articulatory prediction in speaker-dependent and speaker-independent setups, as well as in distinct speech corpora. Additionally, it delve into the impact of weight initialization strategies and the critical role of the one-dimensional convolutional layer in smoothing articulatory features. 
\subsection{Batch Size}
Initially investigated the ideal batch size for this model within the contexts of both speaker-dependent (SD) and speaker-independent (SI) scenarios. The comprehensive result summarized in Table \ref{tab:batch} 

\begin{table}[h!]
 \caption{Performance of Batch size 0,8,16 and 32}
\begin{tabular}{|l |l | l | l | l | l |}
    \hline
    \small{Approach} & \small{Matrix} & \small{B\_0} & \small{B\_8} & \small{B\_16} & \small{B\_32} \\
    \hline
    \multirow{2}{*}{SD} & \small{RMSE} & \small{0.838} & \small{\textbf{0.761}} & \small{0.801} & \small{0.79}\\

     & \small{PCC} & \small{0.791} & \small{\textbf{0.81}} & \small{0.777} & \small{0.774}\\
     \hline
    \multirow{2}{*}{SI} & \small{RMSE} & \small{1.774} & \small{\textbf{1.732}} & \small{1.789} & \small{1.781} \\
     & \small{PCC} & \small{0.632} & \small{\textbf{0.698}} & \small{0.525} & \small{0.522} \\
     \hline

\end{tabular}
\label{tab:batch}
\end{table}

In the case of Batch size 0, no batching was applied, implying that each training example was processed individually without grouping. Among these experiments, Batch size 8 stood out as a noteworthy configuration. It achieved convergence in 20 epochs, resulting in a Validation set RMSE of 0.761 and an impressive PCC of 0.81 in the speaker-dependent (SD) scenario. Same as, It achieved convergence in 23 epochs, resulting in a RMSE of 1.732 and an impressive PCC of 0.698 in the speaker-dependent (SI) mode.The effect of batch size is notably conspicuous in the speaker-dependent (SD) scenario. This could be attributed to the fact that in SD scenarios, the dataset typically comprises samples from a single speaker, which tends to exhibit more consistent patterns. Smaller batch sizes like Batch size 8 allow the model to learn these patterns efficiently. Conversely, in SI scenarios, the dataset contains samples from multiple speakers, leading to greater data variability. This increased variability makes it more challenging for the model to benefit significantly from variations in batch size, which might explain the lesser impact of batch size choice in SI scenarios. Considering the slightly better performance of Batch 8 in both SD and SI scenarios, this work adopts Batch size 8 as the optimal batch size.

\subsection{Speaker Dependent and Speaker Independent}
\begin{figure}[h!]
    \centering
    \includegraphics[width=\columnwidth]{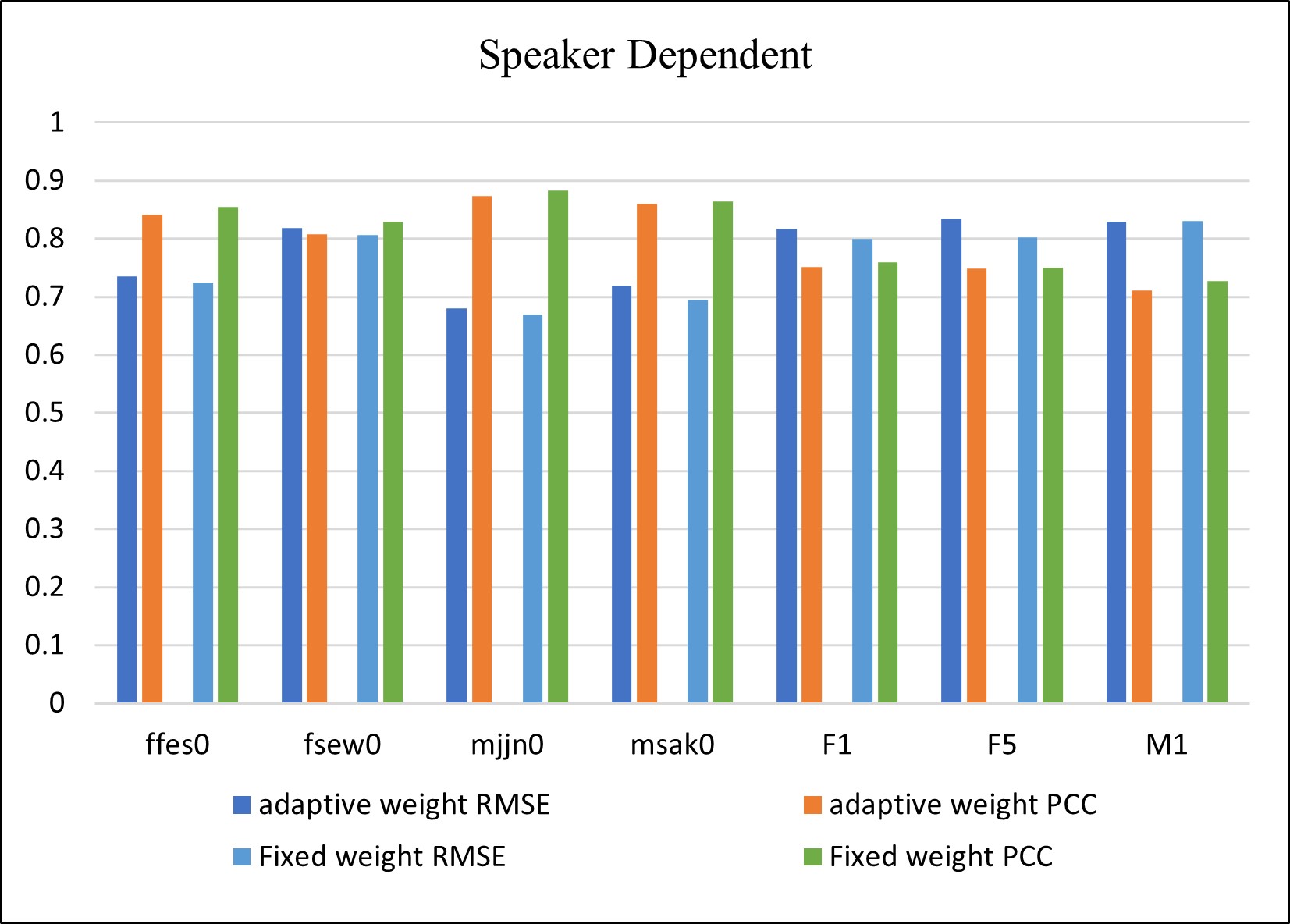}
    \caption{The evaluation of the model in SD approach}
    \label{fig:SD}
\end{figure}

The articulatory properties of each speaker are influenced by individual anatomical and physiological differences, speaking styles, and vocal tract characteristics. These distinctions among speakers lead to variations in the acoustic-to-articulatory mapping. In SD scenarios, where the model is exclusively trained and tested on data from a single speaker, the model can closely align its predictions with the unique characteristics of that speaker's articulatory movements. This focused approach allowed the model to adapt and fine-tune its predictions to the specific articulatory patterns of that speaker. Consequently, the model achieved relatively higher performance metrics. Adaptive weight initialization yields an average RMSE of 0.776 and a PCC of 0.799, while fixed weight initialization results in an RMSE of 0.761 and a slightly higher PCC of 0.810. The detailed performance of each speaker is given in Fig. \ref{fig:SD}

Rather than the adaptive weight, the fixed weight initialization method provided a more consistent performance across speakers. This uniformity suggested that the fixed weights offered a consistent predictive accuracy.  

\begin{figure}[h!]
    \centering
    \includegraphics[width=\columnwidth]{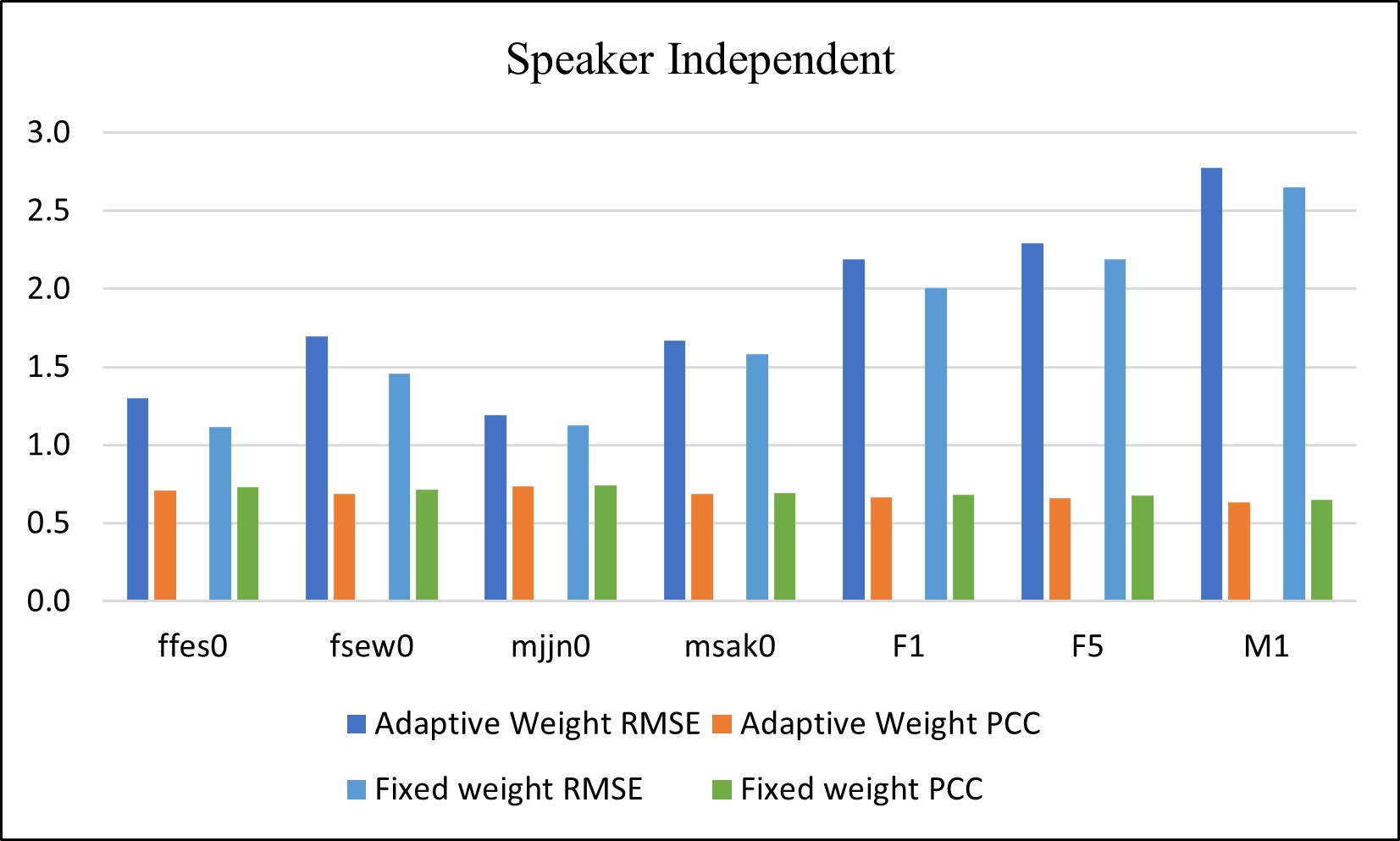}
    \caption{Performance evaluation of each speaker in  SI approach}
    \label{fig:SI}
\end{figure}

The SI scenario combines data from multiple speakers, leading to increased variability in the acoustic features and articulatory trajectories. The SI approach evaluate the generalization feature of the proposed architecture. Adaptive weight initialization results in an average RMSE of 1.873 with a PCC of 0.683, while fixed weight initialization yields a slightly improved average RMSE of 1.732 and a PCC of 0.698. This indicates that the fixed weights provide a more consistent starting point for training and yield slightly better predictive accuracy, on average, across the various speakers in the SI scenario. The performance of the model in each speaker test set is given in Fig.\ref{fig:SI}

The SI scenario introduces a higher degree of complexity and variability compared to the speaker-dependent (SD) scenario. However, the fixed weight initialization offers a modest improvement, showcasing its potential to provide a more stable foundation for handling the challenges posed by SI data.

\begin{table}[h!]
 \caption{SD model comparison with existing BiLSTM with similar data}
\begin{tabular}{|l | l | l |}
\hline
Speaker Dependent & RMSE (mm) & PCC \\
\hline
BiLSTM [39] & 1.429 & 0.753 \\
\hline
BiLSTM [31] & 1.178 & 0.836 \\
\hline
 BiLSTM-CNN & 0.761 & 0.810 \\
\hline
\end{tabular}
\label{tab:SDcomp}
\end{table}

\begin{table}[h!]
\caption{SI model comparison with existing BiLSTM with similar data}
\begin{tabular}{|l | l | l |}
\hline
Speaker Independent & RMSE (mm) & PCC \\
\hline
BiLSTM [39] & 2.247 & 0.308 \\
\hline
 BiLSTM-CNN & 1.732 & 0.698 \\
\hline

\end{tabular}
\label{tab:SIcomp}
\end{table}

The proposed BiLSTM-CNN model compared with the existing BiLSTM models which uses similar dataset for their training and testing. The comparitive study is listed in Table \ref{tab:SDcomp} and Table \ref{tab:SIcomp}. The proposed architecture outperformed than the other model both in SD and SI. 

\subsection{Corpus Dependent and Cross Corpus}
The corpus-dependent evaluation of the model involves assessing its performance on datasets that are specific to particular speech corpora, namely MOCHA and USC. Corpus-dependent evaluation determines the model's robustness across different linguistic contexts and speech databases. the distinction in the linguistic and phonetic parameters of the MOCHA (UK native speakers) and USC (US native speakers) corpora due to the geographical origin of the speakers highlights the importance of conducting corpus-dependent evaluations. Native speakers from the UK and the US often exhibit differences in pronunciation, accent, and speech patterns. These distinctions can manifest in various linguistic aspects, including vowel and consonant articulation, and intonation. 

\begin{figure}[h!]
    \centering
    \includegraphics[width=\columnwidth]{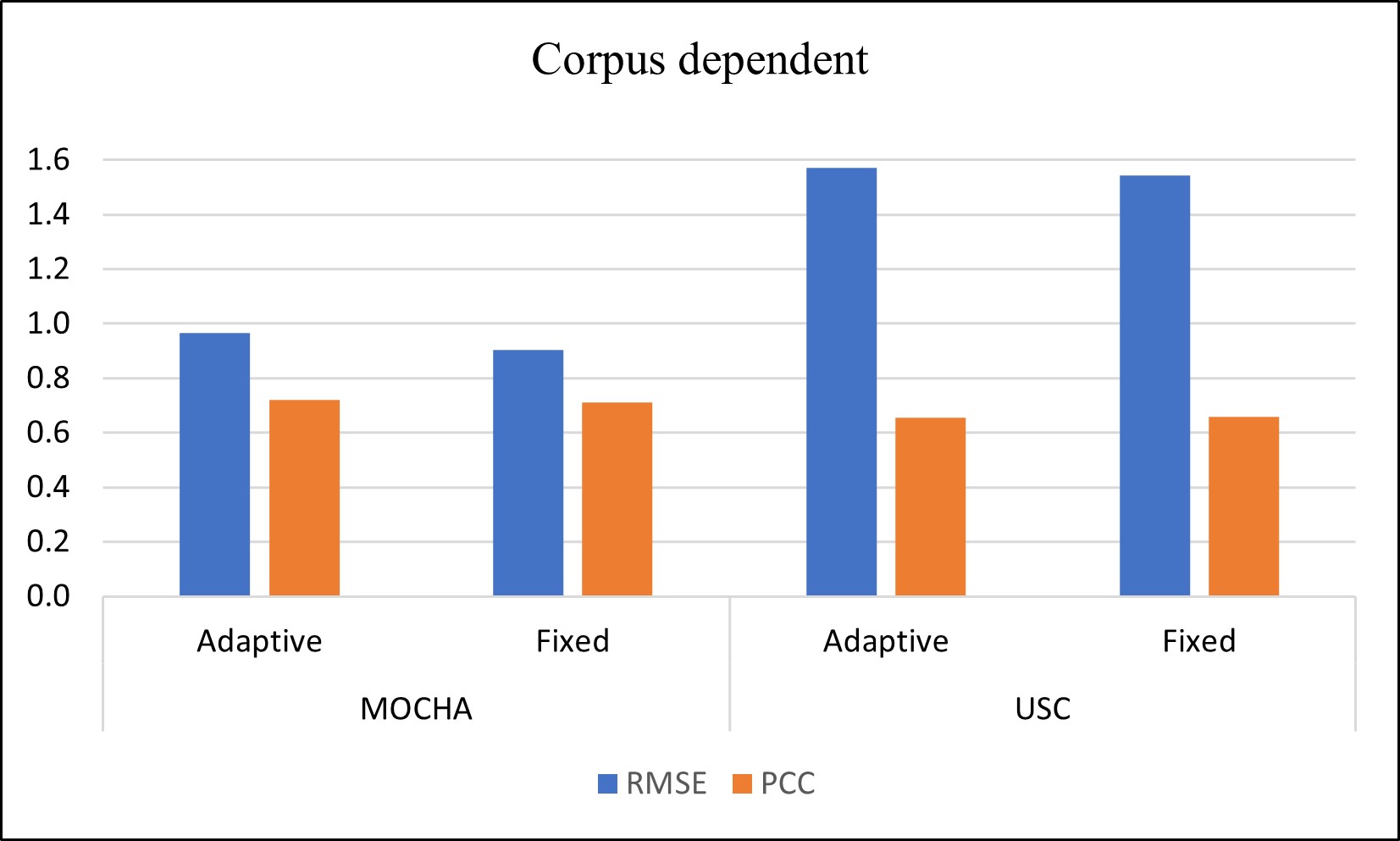}
    \caption{Performance evaluation of each corpus}
    \label{fig:CD}
\end{figure}

For the adaptive weight initialization approach, the model achieved an average RMSE of 0.965, indicating the closeness of its predicted articulatory trajectories to the ground truth. Moreover, the model exhibited a PCC of 0.711, signifying a moderate positive linear relationship between its predictions and the actual articulatory data. Conversely, when employing fixed weight initialization, the model displayed a slightly improved average RMSE of 0.904, suggesting even closer agreement with the ground truth data. Additionally, the PCC increased to 0.721, indicating a slightly stronger positive linear relationship between the model's predictions and the actual articulatory trajectories. In USC corpus, the model achieved an average RMSE of 1.570 and PCC of 0.655 in adaptive weight initialization and in fixed weight initialization the model shown an average RMSE of 1.543 and PCC of 0.659 (Fig. \ref{fig:CD}). 

The model demonstrates promising performance, and the use of fixed weight initialization consistently improves its results in both corpora. This underscores the model's adaptability to different linguistic contexts while emphasizing the potential benefits of fixed weight initialization for enhanced performance.

\begin{figure}[h!]
    \centering
    \includegraphics[width=\columnwidth]{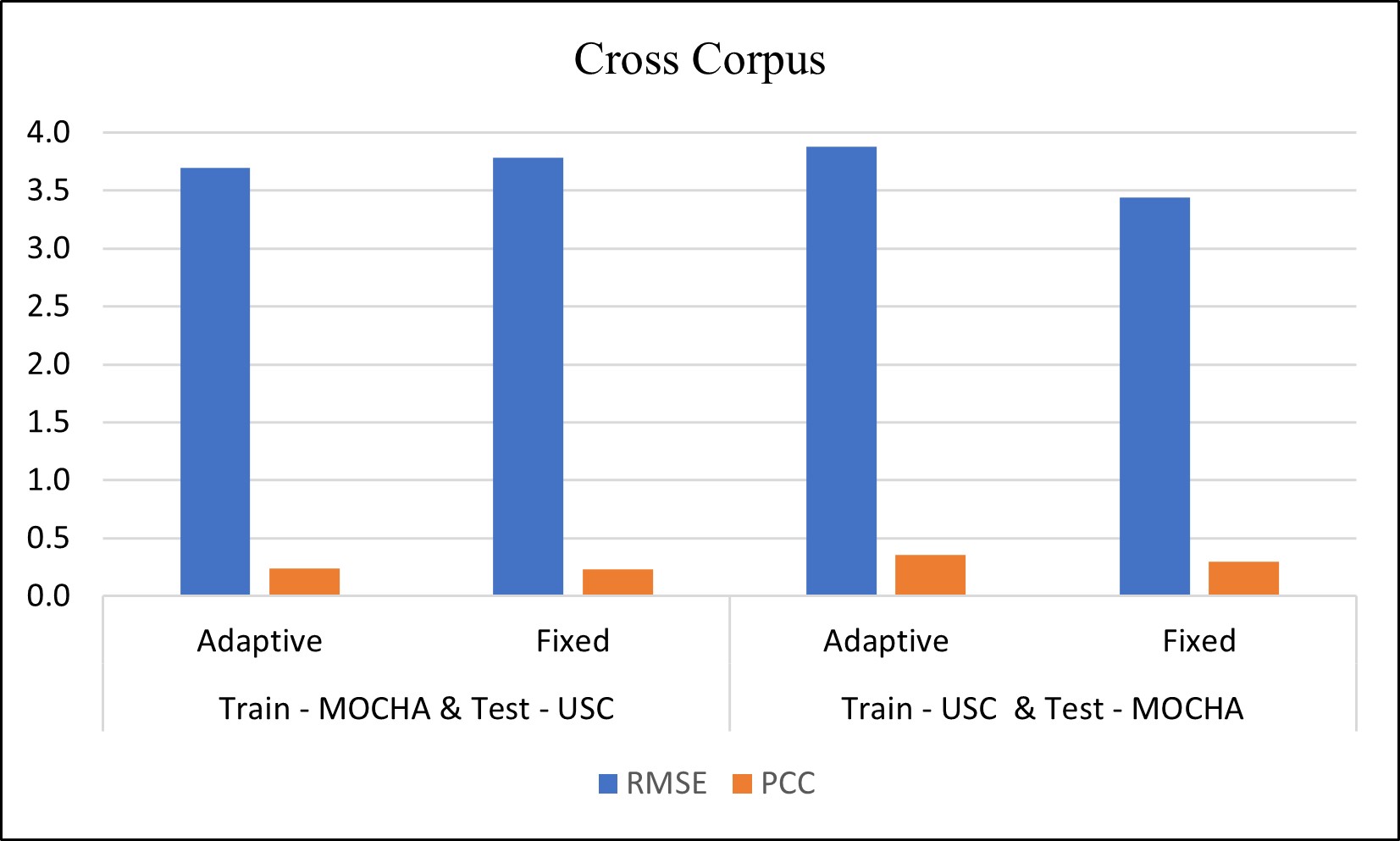}
    \caption{Cross corpus evaluatuion}
    \label{fig:CC}
\end{figure}

When the model is trained on MOCHA, which consists of UK English speakers, and tested on USC, which comprises US English speakers, it encounters a significant challenge in adapting to the phonetic and linguistic differences between the two corpora. This mismatch leads to relatively high RMSE of 3.69, indicating larger prediction errors in articulatory trajectories. Additionally, the PCC is notably low at 0.24, suggesting a weaker linear relationship between predicted and ground-truth trajectories. Conversely, when the model is trained on USC (US English speakers) and tested on MOCHA (UK English speakers), a similar challenge arises. The RMSE remains high at 3.881, indicating substantial prediction errors, while the PCC is 0.353, indicating a somewhat improved linear relationship compared to the previous scenario. The cross-corpus evaluation results, it appears that the choice of weight initialization method (fixed or adaptive) does not significantly impact the model's ability to generalize across different corpora (MOCHA and USC). Both weight initialization methods exhibit similar trends in performance across these datasets. The detailed illustration is presented in Fig. \ref{fig:CC}   

In this comprehensive series of experiments, the primary goal was to assess the performance and robustness of a novel BiLSTM-CNN neural network architecture with fixed weight initialization for smoothing. The experiments delved into several critical aspects of the model's behavior, including weight initialization strategies, the influence of batch size, and the model's capacity to generalize across different speech corpora. The fixed weight initialization has shown better performance in those scenarios where test set have linguistic or phonetic relation with the training set, such as speaker dependent, corpus dependent and to some extend speaker independent. The fixed weight is applied in the 1D convolution for smoothing the predicted articulatory. Smoothing plays a crucial role in articulatory prediction because smoothing contributes to the temporal consistency of articulatory trajectories.
\begin{figure*}[htb!]
    \centering
    \begin{minipage}{0.49\textwidth}
    \centering
    \includegraphics[width=1\textwidth]{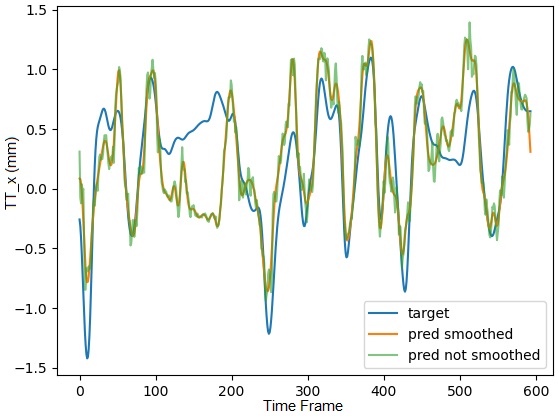}
    \end{minipage}
    \hfill
    \begin{minipage}{0.49\textwidth}
    \centering
    \includegraphics[width=1\textwidth]{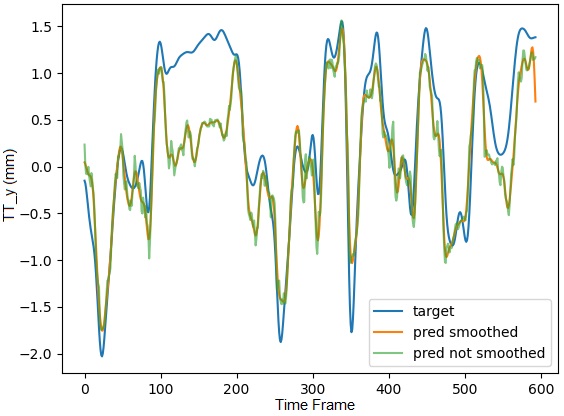}
    \end{minipage} \\
    \begin{minipage}{0.49\textwidth}
    \centering
    \includegraphics[width=1\textwidth]{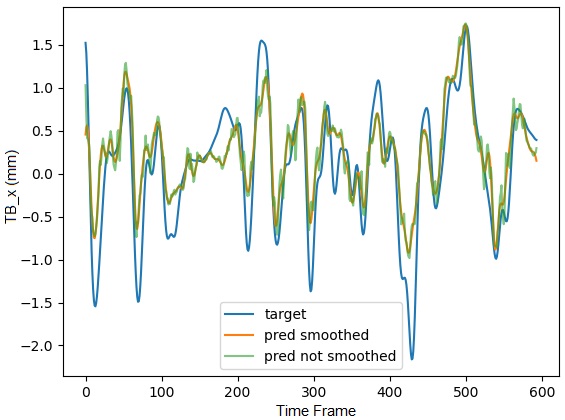}
    \end{minipage}
    \hfill
    \begin{minipage}{0.49\textwidth}
    \centering
    \includegraphics[width=1\textwidth]{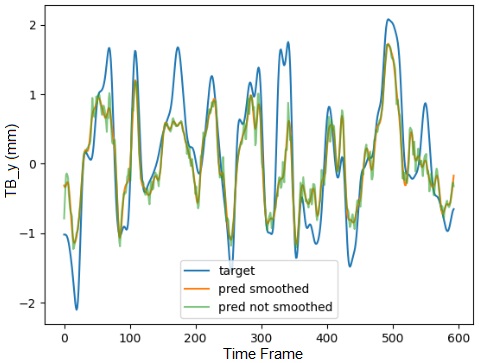}
    \end{minipage}\\
    \begin{minipage}{0.49\textwidth}
    \centering
    \includegraphics[width=1\textwidth]{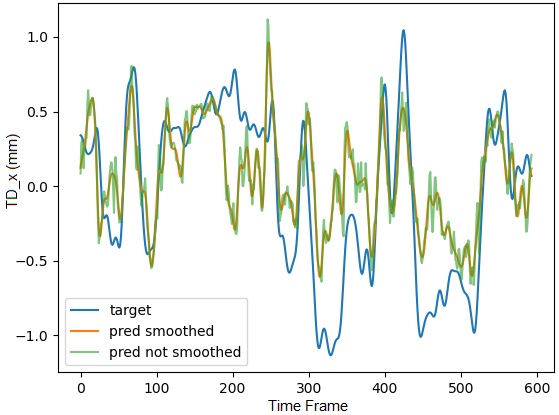}
    \end{minipage}
    \hfill
    \begin{minipage}{0.49\textwidth}
    \centering
    \includegraphics[width=1\textwidth]{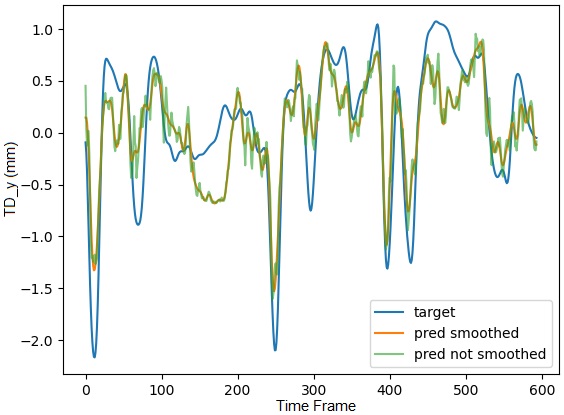}
    \end{minipage}
    \label{fig:ton}
    \caption{Predicted tongue articulatory of the sentence ''Each untimely income loss coincided with the breakdown of a heating system part''}
    \label{fig:AAI}
\end{figure*}

In the natural process of speech production, articulatory movements exhibit a smooth and sequential pattern. Abrupt and jagged changes in predicted articulatory features do not align with this inherent smoothness, making smoothing essential for generating more realistic and linguistically meaningful results. The predicted tongue movements for the production of the sentence \textit{''Each untimely income loss coincided with the breakdown of a heating system part''} using proposed BiLSTM–CNN model is illustrated in Fig. \ref{fig:AAI}.

The target trajectory movements (blue) are continuous and smooth. However, the predicted trajectory may have a jagged representation (green). The predicted trajectory without smoothing and with smoothing has shown in the figure . The smoothed trajectory (red) using one dimensional CNN is correlated closer to the target trajectory. The post processing smoothing reference for comparison study was available on MOCHA dataset. Therefore, the comparison of proposed CNN smoothing with fixed weight initialization in MOCHA corpus based on the Correlation coefficient is presented in Table \ref{tab:FW}.

\begin{table}
    \centering
    \caption{Comparison of Proposed Fixed weight smoothing with other approaches}
    \begin{tabular}{| l | l | l |}
    \hline
    Model& Smoothing Approach& PCC \\
    \hline
     & Direct & 0.816 \\
     CNN[42]& Kalman & 0.827 \\
    & MLPG & 0.841 \\
    \hline
    BiLSTM-CNN & CNN with fixed
    weight & \textbf{0.858} \\
\hline
\end{tabular}
\label{tab:FW}
\end{table}

The contribution of this work is a deep neural network architecture designed with BiLSTM for training and a CNN with fixed weight approach for smoothing the predicted trajectory from the BiLSTM. The comparison illustrated that the proposed BiLSTM-CNN model with fixed weight initialization outperformed the other models, achieving better correlation coefficient than other models in MOCHA corpus. This result signifies that this model provides the most accurate and smooth articulatory predictions even in speaker independent approach.

\section{Conclusion}
This work addresses the challenges in tracking articulatory movements, focusing on the intricate dynamics of the tongue and lips during speech production. To enhance the accuracy of this tracking process, an innovative stacked BiLSTM-CNN architecture is introduced. To evaluate the model's robustness and generalization capabilities, experiments are conducted using two distinct speech datasets, each introducing variations in terms of geographical origin, linguistic characteristics, phonetic diversity, and recording equipment. Both datasets used in this study consist of parallel recordings of speech and corresponding EMA data. This parallel data allows for a robust analysis of articulatory movements during speech production, training and evaluation of predicted articulatory features from acoustic input. The evaluation encompasses various modes, including speaker-dependent and speaker-independent scenarios, as well as corpus-dependent evaluations and conducted cross-corpus assessments, putting the model's adaptability to the test in different linguistic contexts and data sources. One common issue in estimating articulatory features is the presence of jagged, unrefined values, which can hamper the practicality of the results. To rectify this, a one-dimensional CNN filter is employed to effectively smoothen out the estimated articulatory trajectories. A noteworthy approach is the utilization of a fixed weight initialization for the smoothing layer, determined to be the most effective during training, resulting in improved accuracy overall. The cross-corpus evaluation results highlight the difficulties in achieving cross-corpus generalization. The differences in linguistic and phonetic parameters between UK and US English, as well as potential variations in speaking styles and data collection methods, contribute to the observed performance disparities. To enhance cross-corpus performance, domain adaptation techniques or corpus-specific model fine-tuning will be explored.
\section*{Acknowledgement}
We extend our heartfelt gratitude to Ms. Manju S, audiologist and speech language pathologist at National Institute for Speech and Hearing (NISH) for her invaluable contributions, generous sharing of expertise and unwavering encouragement throughout this work.  

\bibliography{acl_latex}

\end{document}